\documentclass{article}
\usepackage{graphicx}
\usepackage[totalwidth=450pt,totalheight=650pt,a4paper]{geometry}
\newcommand{\et}{\sl et al. \rm}

\title{Wide Field Surveys and Astronomical Discovery Space}

\author{A.Lawrence \\
Institute for Astronomy, SUPA\footnote{Scottish Universities Physics Alliance}, University of Edinburgh, \\
Royal Observatory, 
Blackford Hill, Edinburgh EH9 3HJ }

\date{A review for publication in {\it Astronomy and Geophysics} \\ Feb 27th 2007}

\begin{document}

\maketitle

\baselineskip 12pt 


\begin{abstract}

I review the status of science with wide field surveys. For many decades surveys have been the backbone of astronomy, and the main engine of discovery, as we have mapped the sky at every possible wavelength. Surveys are an efficient use of resources. They are important as a fundamental resource; to map intrinsically large structures; to gain the necessary statistics to address some problems; and to find very rare objects. I summarise  major recent wide field surveys - 2MASS,  SDSS, 2dfGRS, and UKIDSS - and look at examples of the exciting science they have produced, covering the structure of the Milky Way, the measurement of cosmological parameters, the creation of a new field studying substellar objects, and the ionisation history of the Universe. I then look briefly at upcoming projects in the optical-IR survey arena - VISTA, PanSTARRS, WISE, and LSST. Finally I ask, now we have opened up essentially all wavelength windows, whether the exploration of survey discovery space is ended. I examine other possible axes of discovery space, and find them mostly to be too expensive to explore or otherwise unfruitful, with two exceptions : the first is the time axis, which we have only just begun to explore properly; and the second is the possibility of neutrino astrophysics.

\end{abstract}



\section{Why are wide field surveys important ?}  \label{why}

Some astronomical experiments are direct, in that measurements are made of some piece of sky, and these measurements are then used for a specific scientific analysis. The essence of a survey however is that extracting science is a {\em two step process}. First we summarise the sky, usually by taking an image and then running pattern recognition software to produce a catalogue of objects each with a set of measured parameters. When this summary is made, we can then do the science with the catalogue; the archive becomes the sky. There are many such archives, distributed around the world in online structured databases; querying such databases is a growing mode of scientific analysis. This of course is why survey databases have played such a central role in the worldwide {\em Virtual Observatory} initiatives.  

Why is this two-step process a good thing to do ? Firstly, it is {\em cost effective}, because we can perform many experiments using the same data. Secondly, surveys are a {\em resource} that can support other experiments. This can mean for example creating samples of objects which are `followed up', i.e. observed in detail, on other facilities (eg getting spectra of galaxy samples). Conversely, interesting objects discovered by other experiments can be matched against objects in the standard survey catalogues, so that one quickly has the optical flux of a new gamma-ray source. (Should this be called follow-down ?). Finally, surveying the sky can produce {\em surprises}. First looks in new corners of parameter space have often discovered completely new populations of objects. Historically, surveys have been the main {\em engine of discovery} for astronomy.

Why are {\em wide angle} surveys important, as opposed to the deepest possible pencil beams ? The key point here is that in Euclidean space, time spent surveying more area increases volume much faster than time spent going deeper. (The argument that wide angles produce large samples faster breaks down when the differential source count slope is flatter than 1, which for example occurs for galaxies fainter than about B$\sim 23$. Also of course, sometimes, one simply has to go deep, for example to survey at some given large redshift.) Many astronomical problems need large samples of objects to address them. Sometimes this is because one wants accurate function estimation -- for example to test theories of structure formation, one wants to estimate the galaxy clustering power spectrum to an accuracy of around 5\% in many bins over a wide range of scale. Sometimes large samples are needed to recover a very weak signal from noise -- for example the net alignment of many random galaxy ellipticities produced by weak lensing by intervening dark matter.  The second reason for maximising volume as quickly as possible is to find {\em rare objects}, such as the hoped for Y dwarfs and $z=7$ quasars; to a given depth there might be only a handful over the whole sky. Finally, some objects of astronomical study simply have intrinsically large angular scale - for example the Milky Way, the galaxy clustering dipole, or open clusters of stars, which can be tens of degrees across.

\section{Major surveys}  \label{surveys}

Surveys are the core of astronomy. This has always been true of course, from Ptolemy through the New General Catalogue, to the Carte du Ciel, but it has been certainly been the case in the last few decades. Table 1 lists some of the best known imaging surveys in each wavelength regime. (I have also included a few redshift surveys as a distinct set). This is only a selection, and is biased towards my own favourites, so apologies to those whose own surveys aren't listed. The point to note is that these names are as immediately recogniseable to every astronomer as are the names of famous telescopes and satellites - Palomar, AAT, Ariel-V, etc. The data in these catalogues are of everyday use and have been the source of many discoveries. Many of the older surveys were classic examples of opening a completely new window on the Universe - 3C, IRAS, and 3U in particular, though I think it is also fair to include the CfA redshift survey in this category, as it gave us the first real feel for the three dimensional structure of the Universe, with bubbles, filaments, and walls.  The 1-XMM catalogue is slightly different, in that it wasn't planned as a coherent single survey, but is the uniformly processed summation of XMM pointings over the sky.

\begin{table}
\centering
\caption{\it Examples of major astronomical surveys from recent decades}
\label{examples}
\begin{tabular}{ll}\\ 
\bf Type \rm  & \bf Survey Examples \rm \\
              & \\
\hline
              & \\
Radio         & 3C, PKS, 4C, FIRST \\
IR	          &	IRAS-PSC, ELAIS, 2MASS, UKIDSS \\
Optical       &	APM, SuperCOSMOS, SDSS, CFHTLS \\
X-ray         & 3U, 2A, HEAO-A, 1-XMM \\
$z$-surveys   & CfA-z, QDOT, 2dFGRS, SDSS-z \\ 
              & \\
\hline
\end{tabular}
\end{table}


%

Over the last 5-10 years the most important major new surveys have been in the optical-IR - 2MASS, SDSS, 2dFGRS, and now UKIDSS, which started in 2005. I will summarise each of these briefly in turn. Some highlight science results are in the next section.

{\bf The Two Micron All Sky Survey : 2MASS.} 

2MASS broke new ground, as it was the first real sky survey at near infa-red wavelengths. At near-IR wavelengths we see roughly the same Universe as in the visible light regime, but with some key improvements. Extinction is much less; we can see pretty much clean through the Milky Way, and can find reddened versions of objects such as quasars. Cooler objects such as brown dwarfs can be found, with the most extreme objects essentially invisible in standard optical bands. Cleaner galaxy samples can be constructed, with high redshift objects easier to find. Colour combinations with optical bands have proved especially good at finding rare objects, such as the new T-dwarf class of brown dwarfs.

2MASS used two dedicated 1.3m telescopes, in Mt Hopkins, Arizona, and CTIO, Chile. Each telescope was equipped with a three-channel camera, each channel consisting of a 256$\times$256 HgCdTe array, so that observations could be made simultaneously at J (1.25 microns), H (1.65 microns), and K$_s$ (2.17 microns). One interesting innovation was the use of large pixels, maximising survey speed, requiring micro-stepping to improve sampling. The survey started in June 1997 June and completed in February 2001. The full data release occurred in March 2003, including both an Atlas of images and a catalogue of almost half a billion sources. To a point source limit of 10$\sigma$, the catalogue depth is J=16 H=15 K$_s$=14.7, almost five orders of magnitude deeper than any comparable IR survey. However, for the colours of many astronomical objects, this is still two orders of magnitude shallower than modern optical surveys. 


\begin{figure}

\centerline{\includegraphics[width=120mm,angle=0]{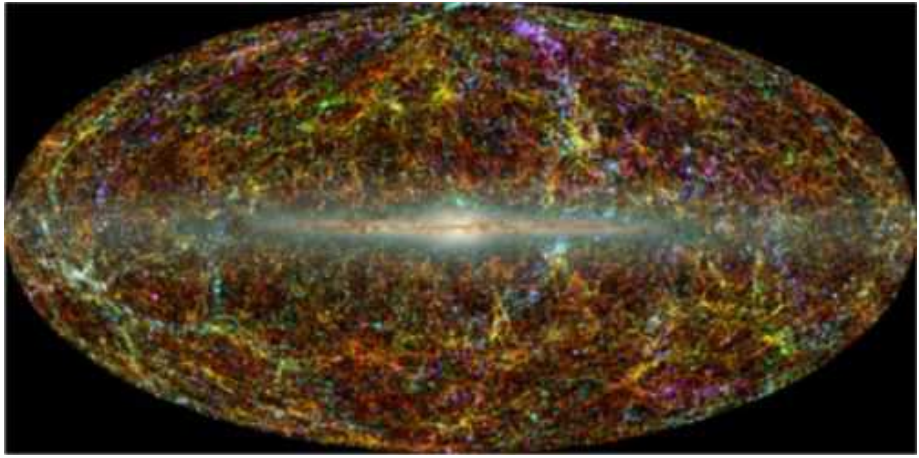}}

\caption{\it All sky distribution of 2MASS catalogues. Point sources are shown as white dots. Extended sources are coloured according to estimated redshift, based either on known values, or estimated from K magnitude. Blue are the nearest sources (z $<$ 0.01); green are at moderate distances (0.01 $<$ z $<$ 0.04) and red are the most distant sources that 2MASS resolves (0.04 $<$ z $<$ 0.1). Taken from Jarrett (2004). }

\end{figure}

%

The core reference for 2MASS is Skrutskie \et 2006). Further information can be found at the IPAC (http://www.ipac.caltech.edu/2mass/) and UMASS(http://pegasus.phast.umass.edu/) sites. Data access is through the IRSA system at http://irsa.ipac.caltech.edu/ .

{\bf The Sloan Digital Sky Survey : SDSS.}  

The SDSS project has produced a survey of 8,000 square degrees of sky at visible light wavelengths, approximately two magnitudes deeper than the historic Schmidt surveys, and in addition has carried out a spectroscopic survey of objects selected from the imaging survey. The project used a dedicated 2.5m telescope at Apache Point, New Mexico, and a camera covering 1.5 square degrees. Survey operations used a novel drift scan approach; the 30 CCDs on the camera are arranged in five rows each sensitive to a separate filter band (u,g,r,i,z); the telescope is parked in a given position and the sky allowed to drift past. The spectroscopic survey is carried out using a 600-fibre system, on separate nights spliced into the imaging programme. This then required the data processing pipeline to keep up in almost real time. Public data access has been announced in a series of staged releases, culminating in June 2006 with DR5, which contains a catalogue of around 200 million objects, and spectra for around a million galaxies, quasars, and stars. An extended programme, cunningly called SDSS-II, has now commenced, and is expected to continue through 2008.

SDSS has been arguably the most successful survey project of recent times, with many hundreds of scientific papers based directly on its data, and having an impact on a very large range of astronomical topics - large scale structure, the highest redshift quasars, the structure of the Milky Way, and many other things besides. This may seem surprising, as visible light sky surveys covering the whole sky have been available for decades, and available as digitised queryable online databases for some years (eg the Digitised Sky Survey (DSS : see http://archive.stsci.edu/dss/) or the SuperCOSMOS Science Archive (SSA : see http://surveys.roe.ac.uk/ssa). There are several reasons for the success of SDSS. The first reason is of course the spectroscopic database, matched only by 2dF (see below). The second reason is the wider wavelength range, with filters carefully chosen and calibrated to optimise various kinds of search. The third reason is the improvement in quality - not only is SDSS a magnitude or two deeper than the Schmidt surveys, but the seeing is markedly better. The fourth reason, shared by 2MASS, is the quality of the online interface - well calibrated, reliable, and documented data were available promptly, and with the ability to do online analysis rather than just downloading data. This has made it easy for astronomers all over the world to jump in and benefit from SDSS.

The core reference for SDSS is York \et 2000). Further information can be found at http://www.sdss.org, which also contains links to data access via SkyServer.

{\bf The UKIRT Infrared Deep Sky Survey : UKIDSS.}   

UKIDSS is the near-infrared equivalent of the SDSS, covering only part of the sky, but many times deeper than 2MASS. The project has been designed and implemented by a private consortium, but on behalf of the whole ESO member community, and after a short delay, the world. It uses the Wide Field Camera (WFCAM) on the UK Infrared Telescope (UKIRT) in Hawaii, and is taking roughly half the UKIRT time over 2005-2012. WFCAM has an instantaneous field of view of 0.21 sq.deg, much larger than any previous large facility IR camera. Put together with a 4m telescope, this makes possible an ambitiuous survey. It is estimated that the effective volume of UKIDSS will be 12 times that of 2MASS, and the effective amount of information collected 70 times larger.  UKIDSS is not a single survey, but a portfolio of five survey components. Three of these are wide shallow surveys, to K$\sim 18-19$, and covering a total of $\sim 7000$ sq.deg - the Galactic Plane Survey (GPS); the Galactic Clusters Survey (GCS); and the high latitude Large Area Survey (LAS). Then there is a Deep Extragalactic Survey (DXS), covering 35 sq.deg to $K\sim 21$, and an Ultra Deep Survey (UDS), covering 0.77 sq.deg. to $K\sim 23$. In all cases, there is the maximum possible overlap with other multiwavelength surveys and key areas, such as SDSS, the Lockman Hole, and the Subaru Deep Field. 

The aim of UKIDSS is to provide a public legacy database, but the design was targeted at some specific goals - for example, to measure the substellar mass function, and its dependence on metallicity; to find quasars at $z=7$; to discover Population II brown dwarfs if they exist; to measure galaxy clustering at $z=1$ and $z=3$ with the same accuracy as at $z=0$; and to determine the epoch of spheroid formation. Like SDSS, data are being released in a series of stages. At each stage the data are public to astronomers in all ESO member states, and world-public eighteen months later. Data are made available through a queryable interface at the WFCAM Science Archive (WSA : http://surveys.roe.ac.uk/wsa). Three data releases have occured so far -- the ``Early Data Release'', and Data Releases One and Two (DR1 and DR2) which contain approximately 10\% of the likely full dataset. 

UKIDSS is summarised in Lawrence \et (2007) , and technical details of the releases are described in Dye \et (2006) and Warren \et (2007). 

{\bf Redshift surveys : 2MRS/6dFGRS; 2dFGRS and SDSS-z.} 

Systematic redshift surveys based on galaxy catalogues from imaging surveys were one of the big success stories of the 1970s--90s, culminating in the all-sky z-survey based on the IRAS galaxies, the PSC-z (Saunders \et 2000). The most ambitious surveys to date have however been carried out over the last five years. The first example is the construction of a complete all-sky redshift survey based on galaxies in the 2MASS Extended Source Catalog (XCS) to a depth of K$_S$=12.2, containing roughly 100,000 galaxies. In the south, observations are carried out at the UK Schmidt, as part of the 6dfGRS project (Jones \et 2004, http://www.aao.gov.au/local/www/6df/); in the North observations are being carried out by a CfA team at Mt Hopkins, Arizona (see http://cfa-www.harvard.edu/$\sim$huchra/2mass/). The survey is part way through, but has already been used to measure the dipole anisotropy of the local universe (Erdodgu \et 2005).

Two very successful projects have completed redshift surveys of smaller area, but reaching considerably deeper, containing hundreds of thousands of galaxies. The first, in the Northern sky, is SDSS-z, the spectroscopic component of SDSS, as described above. The second, in the southern sky, is the 2dF Galaxy Redshift Survey (2dFGRS; Colless \et 2001). This was based on galaxies selected from the APM digitisation of UK Schmidt plates (Maddox \et 199x), and observed using the Two Degree Field (2dF) facility at the Anglo-Australian Telescope, which has 400 independent fibres. The 2dFGRS obtained spectra for 245591 objects, mainly galaxies, brighter than a nominal extinction-corrected magnitude limit of b$_J$=19.45, covering 1500 square degrees in three regions. The final data release was in June 2003. More information, and data access, is available at http://www.mso.anu.edu.au/2dFGRS/.

These two surveys have produced a range of science, but have concentrated on making the best possible measurement of the power spectrum of galaxy clustering, and together with WMAP and supernova programme results, have produced the definitive estimates of the cosmological parameters, leading to the current `concordance cosmology'.

\section{Recent survey science highlights}  \label{science}

I have picked out a handful of results from the optical-IR surveys of the last few years, including the first results from UKIDSS, to illustrate the power of the survey approach.

{\bf Panoramic mapping : the structure of the Milky Way.} 

Two topics which clearly benefit from a map covering 4$\pi$ sr, and with low extinction, are the structure of the Milky Way, and the structure of the local extragalactic universe. Figure 1, taken from Jarrett (2004), illustrates the impact 2MASS has on both these topics, showing both the Point Source Catalog (mostly stars) and the Extended Source Catalog (mostly galaxies at z$<$0.1). For the first time, we can see the Milky Way looking like other external galaxies, with disc, bulge, and dust lane. Some of the most important scientific results however have come from looking at subsets of the stellar population. Figure 2, from Majewski \et (2003), shows the sky distribution of M giants selected from the 2MASS PSC, a selection which traces very large scale structures while removing the dilution of local objects, using a few thousand stars out of the catalogue of half a billion. From APM star counts we already knew of the existence of the Sagittarius dwarf, swallowed by the Milky Way (Ibata, Irwin and Gilmore 1994), but now we can see its complete structure including an extraordinary 150 degree tidal tail. Its orbital plane shows no precession, indicating that the Galactic potential within which it moves is spherical. The Earth is currently close to the debris, which means that some very nearby stars are actually members of the Sagittarius dwarf system. Interestingly, Sagittarius seems to contribute over 75\% of of high latitude halo M giants, with no evidence for M giant tidal debris from the Magellanic clouds.


\begin{figure}

\includegraphics[width=150mm,angle=0]{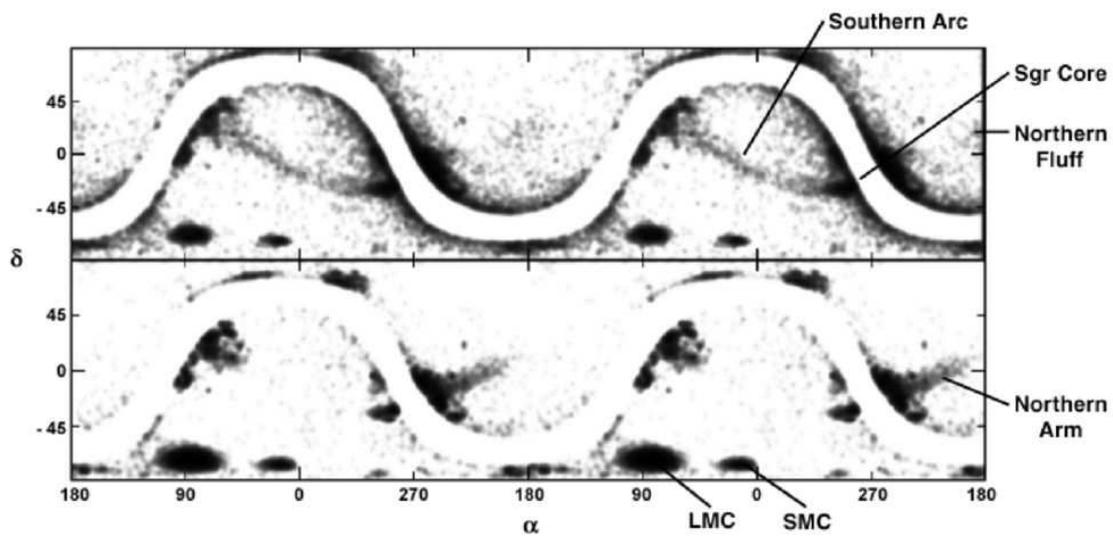}

\caption{\it Smoothed maps of the sky in equatorial coordinates showing the 2MASS point source catalogue optimally filtered to show the Sagittarius dwarf;  southern arc (top), and the Sagittarius dwarf northern arm (bottom). Two cycles around the sky are plotted to demonstrate the continuity of features. The top panel uses $11 < Ks < 12$ and $1.00 < J-Ks < 1.05$. The bottom panel uses $12 < Ks < 13$ and $1.05 < J-Ks < 1.15$. Taken from Majewski et al. (2003). }

\end{figure}

%

SDSS, although not a panoramic survey, has also been very important for Galactic structure and stellar populations, with the five widely spread bands making it possible to derive stellar types and so photometric parallaxes. Juric \et (2005) derive such parallaxes for 48 million stars. They fit a combination of oblate halo, thin disk, thick disk, but also find significant `localised overdensities', including the known Monoceros stream, but also a new enhancement towards Virgo that covers 1000 sq.deg. This then maybe another dwarf galaxy swallowed by the Milky Way.

{\bf Large sample statistics : galaxy clustering and the cosmological parameters.} 


\begin{figure}

\includegraphics[width=160mm,angle=0]{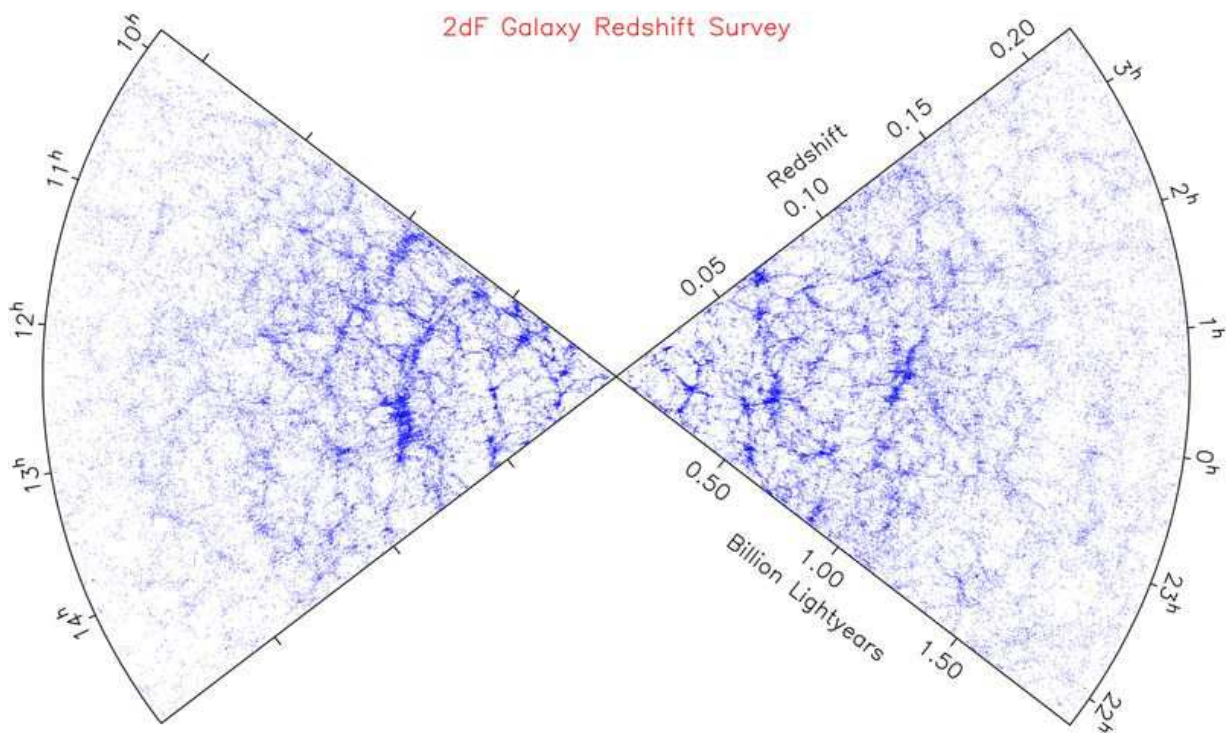}

\caption{\it Cone diagram showing projected distribution of galaxies in 2dFGRS. Taken from Peacock (2002).}

\end{figure}

%

The SDSS-z and 2dFGRS surveys illustrate the power of the survey approach in two ways. First, significant volume is needed to map out large scale structures and overcome shot noise on the largest scales. Figure 3 is a cone diagram for all the 2dFGRS galaxies, showing the richness of structure that is only possible to map out with both a large volume {\em and} density. Second, large numbers are needed to make a good enough estimate of the power spectrum of galaxy clustering. This is illustrated in Fig 4a, which shows the power spectrum derived from 2dFGRS compared to various model predictions (data from Percival \et (2001), figure from Peacock (2002)). To distinguish models with differing matter density in the interesting range requires accuracy of a few percent over a very wide range of scales; to have a chance of measuring small scale features predicted by models including a significant baryon fraction requires many samples across this wide range, with of the order 10$^3$ galaxies per bin to achieve the required accuracy. These wiggles are due to acoustic oscillations in the baryon component of the universe at early times. In the Percival \et paper, only a limit could be placed on these oscillations, but they were statistically detected in the fimal 2dFGRS data (Cole \et 2005). However, in another good example of filtering out a tracer sub-sample from a very large sample, the first baryon peak was much more clearly seen in the correlation function of Luminous Red Galaxies (LRGs) selected from SDSS-z (Eisenstein \et 2005; Huetsi 2005; see Fig 4).

2dFGRS and SDSS-z were the first redshift surveys to have large enough scale and depth to overlap the fluctuation measurements from the CMB, enabling degeneracies in the estimation of cosmological parameters to be broken, and accuracy to be increased by a factor of several. Several key papers made joint analyses of the galaxy and CMB datasets (Percival \et 2002; Efstathiou \et 2002; Tegmark \et 2003; Pope \et 2004) arriving at broadly consistent answers. We now know what kind of universe we live in : a geometrically flat universe dominated by vacuum energy (75\%), with some kind of cold dark matter at about 21\% and ordinary baryons 4\%. 
The equation of state parameter for the dark energy has been limited to $w<-0.52$ (Percival \et 2002), and the total mass of the neutrinos to $m<$1 eV (Tegmark \et 2003; Elgaroy \et 2002)

The Deep eXtragalactic Survey (DXS) of UKIDSS will produce a galaxy survey over a volume as large as that of 2dFGRS or SDSS, but at $z=1$. A redshift survey of this sample is a prime target for future work. 


\begin{figure}

\includegraphics[width=72mm,angle=0]{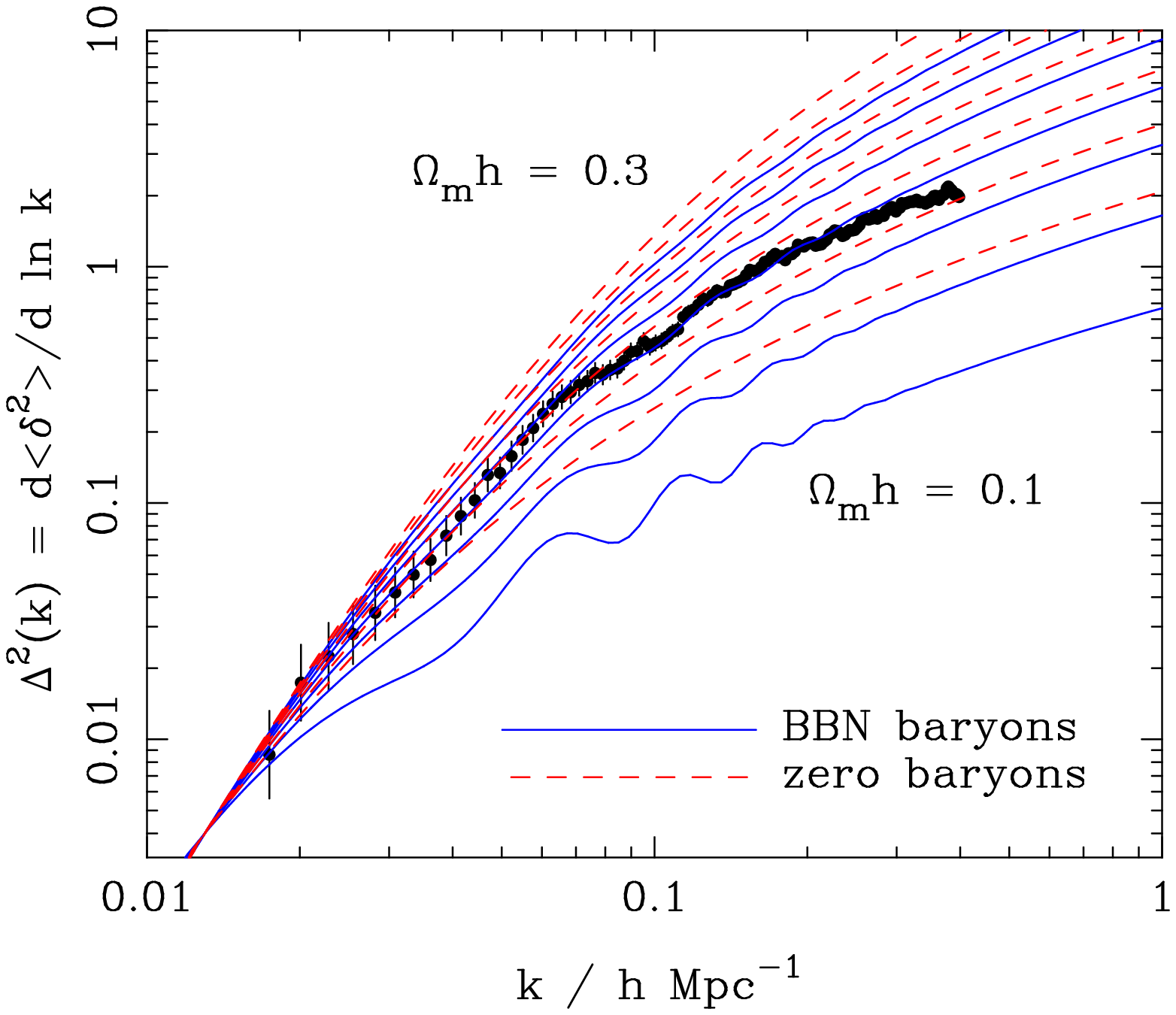}
\includegraphics[width=87mm,angle=0]{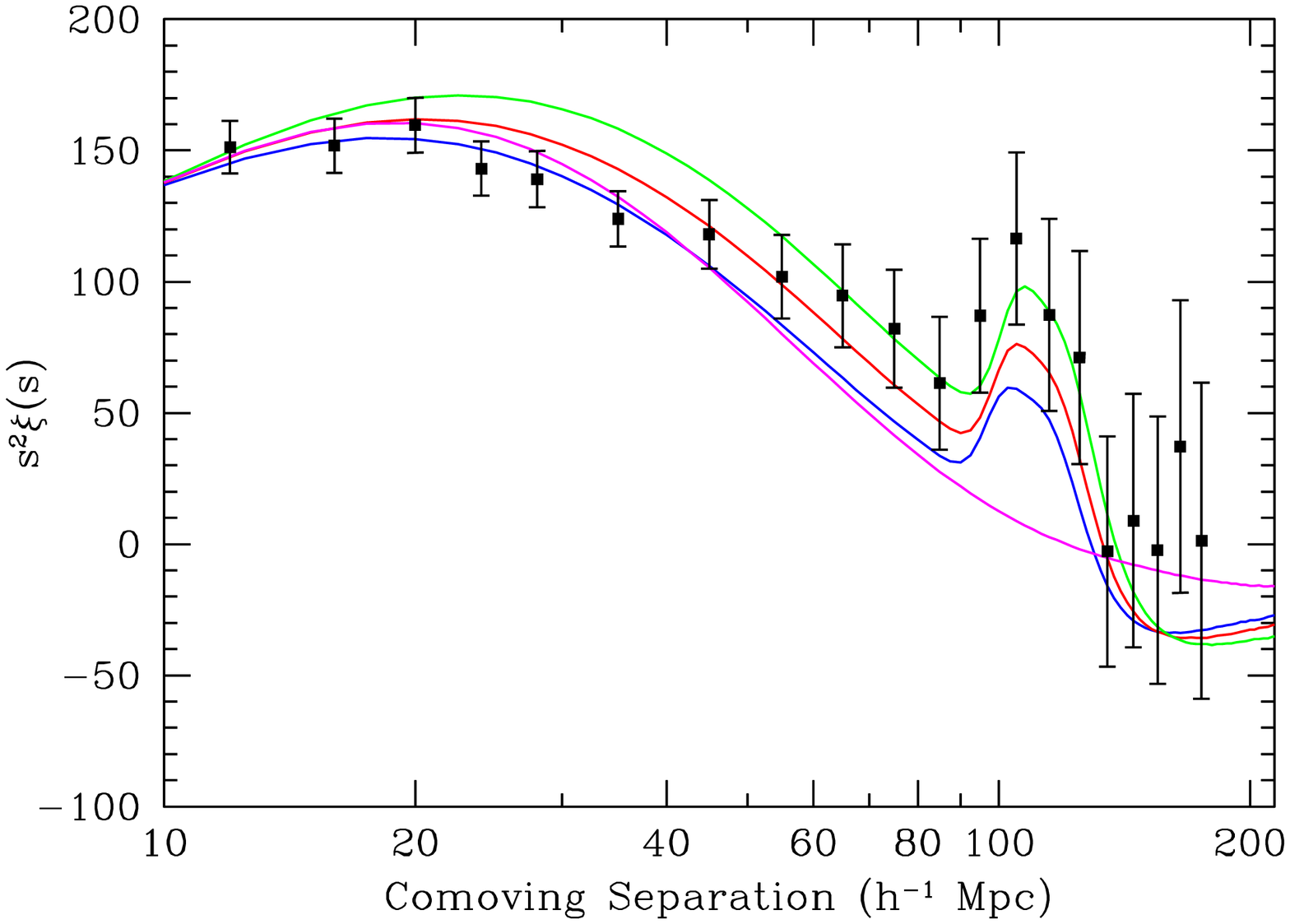}

\caption{\it  (a) Power spectrum from 2dFGRS, compared to various model predictions. Taken from Percival et al. (2001). (b)
Correlation function of Luminous red Galaxies in the SDSS-z sample, showing the first baryon acoustic oscillation peak. Taken from Eisenstein et al. (2005)}

\end{figure}

%

{\bf Rare objects : Brown Dwarfs.} 

Infrared surveys have transformed the study of the substellar regime, blurring our idea of what it means to be a star. For many years, until the first discovery of the very faint IR companion of GL 229 (i.e. GL229B) by Nakajima \et (1995), the possibility of star-like objects which never ignite nuclear burning was only a speculation. Within a year of the start of 2MASS, Kirkpatrick \et (1999) had found 20 brown dwarfs in the field, increasing the number of known brown dwarfs by a factor of four, and had defined two new stellar spectral types - L and T. (These strange designations were determined by the fact that various odd stellar types had already used up nearly all the other letters of the alphabet.) The transition from M to L was defined by the change of key atmospheric spectral features from those of metal oxides to metal hydrides and neutral metals; the transition from L to T by the appearance of molecular features such as methane - as seen in solar system planets.  The effective temperature for L dwarfs is in the range T$\sim$1500 -- 2000 K, and for T-dwarfs T$\sim$1000 --1500 K. As of the time of writing, almost 600 brown dwarfs are known. Most of these are L-dwarfs, but almost 60 T-dwarfs have now been found in a series of 2MASS papers (see Ellis \et 2005 and references therein). 

The much deeper UKIDSS search is expected to make significant further advances in two ways. The first is by pushing to ever cooler and fainter objects, hopefully finding examples of a putative new stellar class labelled `Y dwarfs' (the last useable letter left ... see Hewett \et 2006), finding T-dwarfs further than 10pc, and plausibly finding Population II brown dwarfs if they exist. The second advance expected from UKIDSS is the determination of the substellar mass function, through the Galactic Clusters Survey (GCS), and testing whether it is universal or not. These hopes are already being borne out by early UKIDSS results; Warren \et (2007b) report the discovery of the coolest known star, classified as T8.5; and in early results from the GCS programme, 
Lodieu \et (2007) have found 129 new brown dwarfs in Upper Sco, a significant fraction of all known brown dwarfs, including a dozen below 20 Jupiter masses, finding the mass function in the range 0.3 -- 0.01 solar masses to have a slope of index $\alpha$ = 0.6 $\pm$ 0.1.

{\bf Rare objects : the ionisation history of the Universe.} 


\begin{figure}

\includegraphics[width=70mm,angle=0]{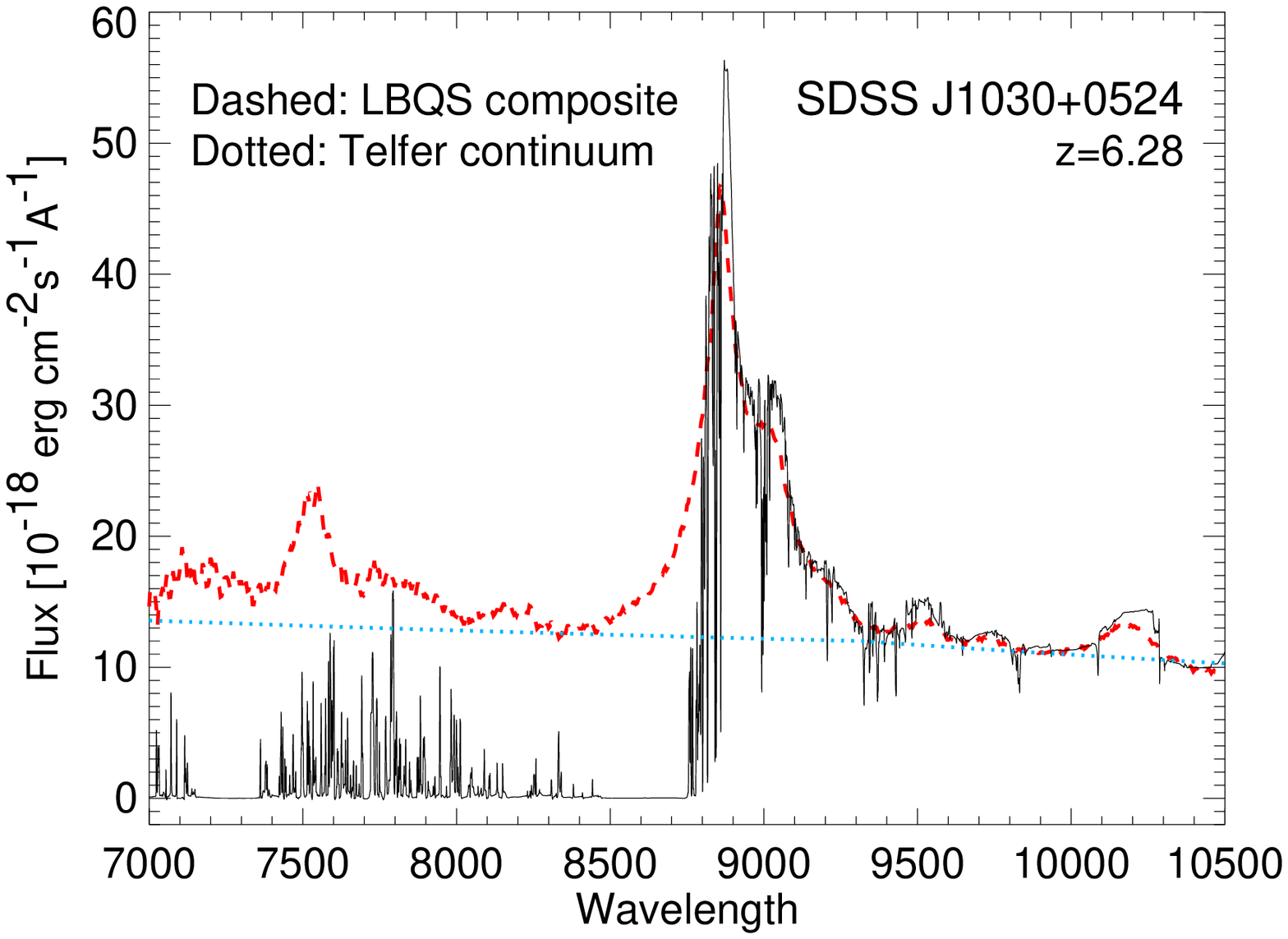}
\includegraphics[width=90mm,angle=0]{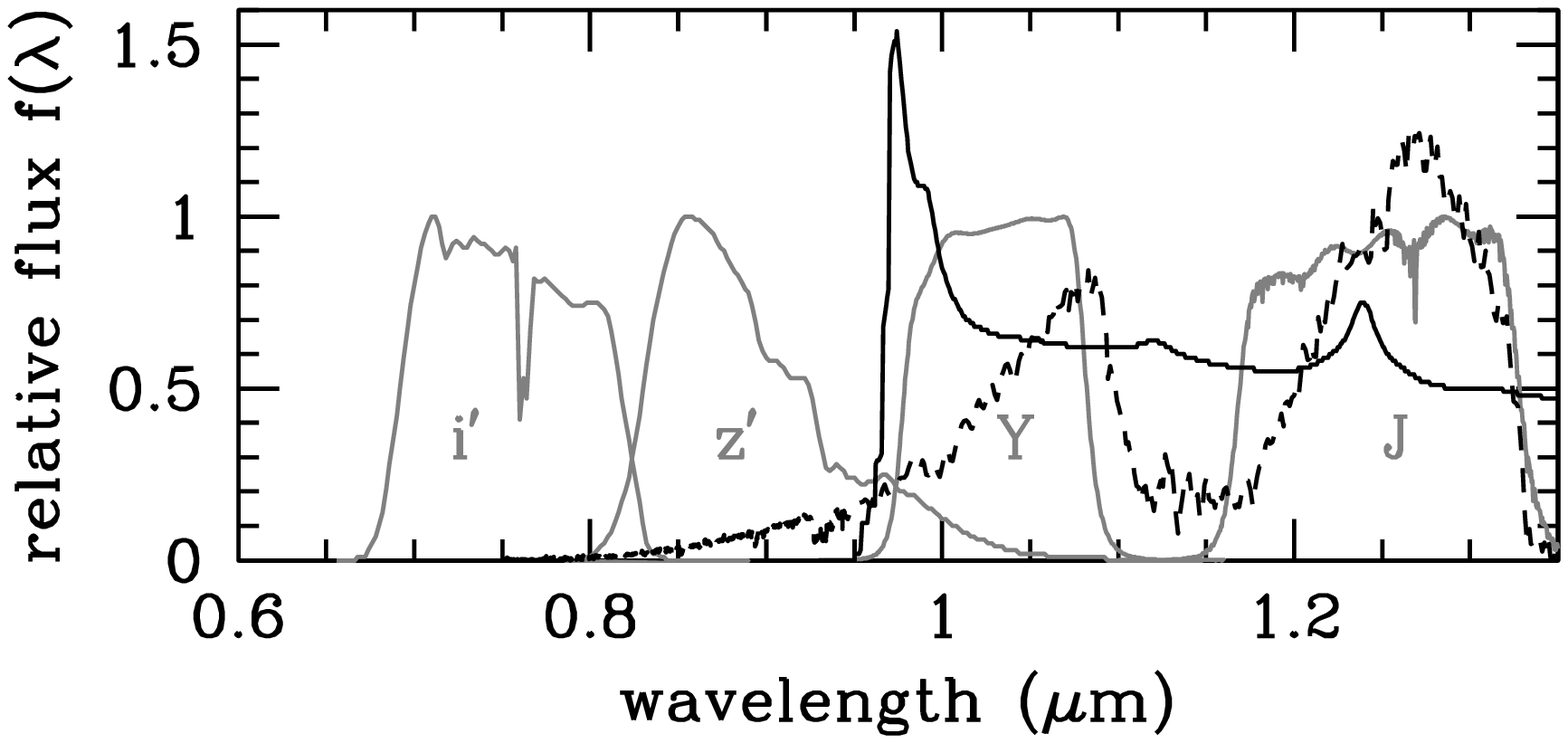}

\caption{\it (a) High resolution spectrum of high redshift quasar found in SDSS. Gunn-Peterson troughs due to Ly$\alpha$ and Ly$\beta$ are the black sections from 8500$\AA$ to 9000$\AA$ and from 7000$\AA$ to 7500$\AA$. Taken from White et al. (2003); original discovery spectrum in Becker et al. (2001). (b) Spectral energy distributions for a $z=7$ quasar and a T-dwarf, compared to filter passbands from SDSS and UKIDSS. Taken from Lawrence et al. (2007).}

\end{figure}

%

An excellent example of the `needle in a haystack' search is looking for very high redshift quasars. Only the most extremely luminous quasars are detectable at these distances, but the space density of such objects is very low; even in a survey with thousands of square degrees there may be only a few present. Luminous and high redshift quasars are interesting for a variety of reasons, but a key target for four decades has been their use as beacons to detect the re-ionisation of the inter-galactic medium. The baryon content of the early universe must have become neutral as it cooled down, but something subsequently re-ionised it, as attempts to find the expected `Gunn-Peterson trough' (Gunn and Peterson 1965) in the spectra of high redshift quasars had failed for many years. This finally changed in 2001 as SDSS  broke the $z=6$ quasar redshift barrier (Fan \et\ 2001) and Becker \et (2001) made the first detection of a Gunn-Peterson trough at $z=6.28$. Figure 5 shows the improved spectrum from White \et (2003).   

Unfortunately this exciting result seemed to conflict with the CMB measurements by the WMAP year-1 data. The degree of scattering required implied that ionisation had already taken place by $z$=11 -- 30 (Kogut \et 2003). Rather than being seen as a contradiction, it seems likely that that re-ionisation was not a single sharp-edged event, but an extended and very likely complex affair, perhaps with multiple stages and even spatial inhomogeneity (see White \et 2003). This opens an entire new field of investigation for understanding the history of the early universe. Rather than a single object locating the transition edge, it is now important to find as many beacons as possible at z$>$6, and to find some beacons in the range $z$ = 7--8. This is one of the key aims of UKIDSS, in combination with SDSS data, looking for z-dropouts. A problem however is that JHK colours of high-z quasars and T-dwarfs become very similar. For this reason, UKIDSS is using a Y-band filter centred at 1.0 $\mu$m. Figure 6 illustrates the point, comparing the spectrum of a quasar redshifted to $z=7$ with that of a T-dwarf brown dwarf.

\section{Next steps in optical-IR surveys}  \label{next}

Three key optical-IR survey projects are to begin soon (VISTA, PanSTARRS, and WISE), with the ultimate in wide-field surveys (LSST) now in the planning stage. Here I briefly summarise each of these. 

{\bf VISTA}.

The Visible and Infrared Survey Telescope for Astronomy (VISTA) is a 4m aperture dedicated survey telescope on Paranal in Chile. It was originally a UK project, aimed ta bothe optical and IR surveys, but became an IR-only ESO telescope during the accession of the UK to ESO. The infrared camera operates at $Z$, $Y$, $J$, $H$, and $K_S$, and contains 16 arrays each of which has 2048$\times$2048 0.33" pixels, covering 0.6 sq.deg. in each shot. VISTA therefore operates in the same parameter space as UKIDSS, but will survey three times faster, and furthermore, 100\% of the time is dedicated to IR surveys. The majority (75\%) of the telescope time is reserved for large public surveys. At the time of writing, these surveys are in the process of final approval, but are likely to include a complete hemisphere survey to $K$=18.5, surveys of the Galactic Bulge and the Milky Way, a thousand sq.deg. survey to $K$=19.5, a 30 sq.deg. survey to $K$=21.5, and a 1 sq.deg. survey to $K$=23. VISTA is expected to begin operations in late 2007. The VISTA web page is at http://www.vista.ac.uk/, and a recent reference is McPherson \et (2006).

{\bf PanSTARRS}.

The power of a survey facility is measured by its {\em \'{e}tendue}, the product of collecting area times field of view. The cost of a telescope, and the difficulty of producing very wide fields, scales steeply with telescope aperture. The idea behind the `Panoramic Survey Telescope and Rapid Response System' (PanSTARRS), a University of Hawaii project, is to produce the maximum  \'{e}tendue per unit cost by building several co-operating wide field telescopes of moderate size. The design has four 1.8m telescopes each with a mosaic array of 64$\times$64 CCD chips covering 7 sq.deg, which will produce an \'{e}tendue an order of magnitude larger than the SDSS facility.
As well as enabling one to produce deep surveys faster, this makes it plausible to cover very large areas of sky repeatedly - thousands of square degrees per night. The prime aim of PanSTARRS is to detect potentially hazardous NEOs, but it will also be used for stellar transits, microlensing studies, and locating distant supernovae to constrain the dark energy problem. The accumulated sky survey will be many times deeper than SDSS, and the expected image quality and stability from Hawaii should allow the best ever mapping of dark matter via weak lensing distortions.

A prototype single PanSTARRS system (`PS1') has recently been built and is being commissioned at the time of writing. The operation and science analysis for PS1 involves an extended `PS1 Science Consortium' with additional partners, from the US, Uk and Germany. Over three years, it is expected to produce a 3$\pi$ steradian survey at $g r i z y$ to $z$=23 with 12 visits, a Medium Deep Survey visiting 12 7 sq.deg. fields with a 4 day cadence, building up a survey to $z$=26, and special stellar transit campaigns and microlensing monitoring of M31. The data will become public at the end of this science programme. 

Information about PanSTARRS can be found at http://pan-starrs.ifa.hawaii.edu/public/

{\bf WISE}. 

The Wide Field Infrared Survey Explorer (WISE) is a NASA MIDEX mission scheduled for launch in 2009 that will fill the gap between UKIDSS/VISTA in the near-IR and IRAS and Akari in the far-IR, surveying the sky in four bands simultaneously (3.3, 4.7, 12, and 23$\mu$m).  The sky survey at 3 and 5$\mu$m is completely new territory; as 12 and 23$\mu$m WISE covers the same territory as IRAS but will be a thousand times deeper. WISE carries a 40cm cooled telescope with a 47 arcmin field of view. It is designed to have a relatively short lifetime - 7 months - but in this time will make a mid-infrared survey of the entire sky in all four bands. WISE will produce significant advances in a number of areas, but especially for objects expected to have temperatures in the hundreds of degrees - the very coolest brown dwarfs, protoplanetary discs, solar system bodies, and obscured quasars. Information about WISE can be found at http://wise.ssl.berkeley.edu/ and in Mainzer \et (2006).

The depths of PS1, UKIDSS-VISTA, and WISE surveys are well matched and will produce a stunning sky survey dataset over a factor of a hundred in wavelength. This is illustrated in Fig. 6, taken from a recent proposal to extend UKIDSS to a complete hemisphere survey.


\begin{figure}

\includegraphics[width=55mm,angle=-90]{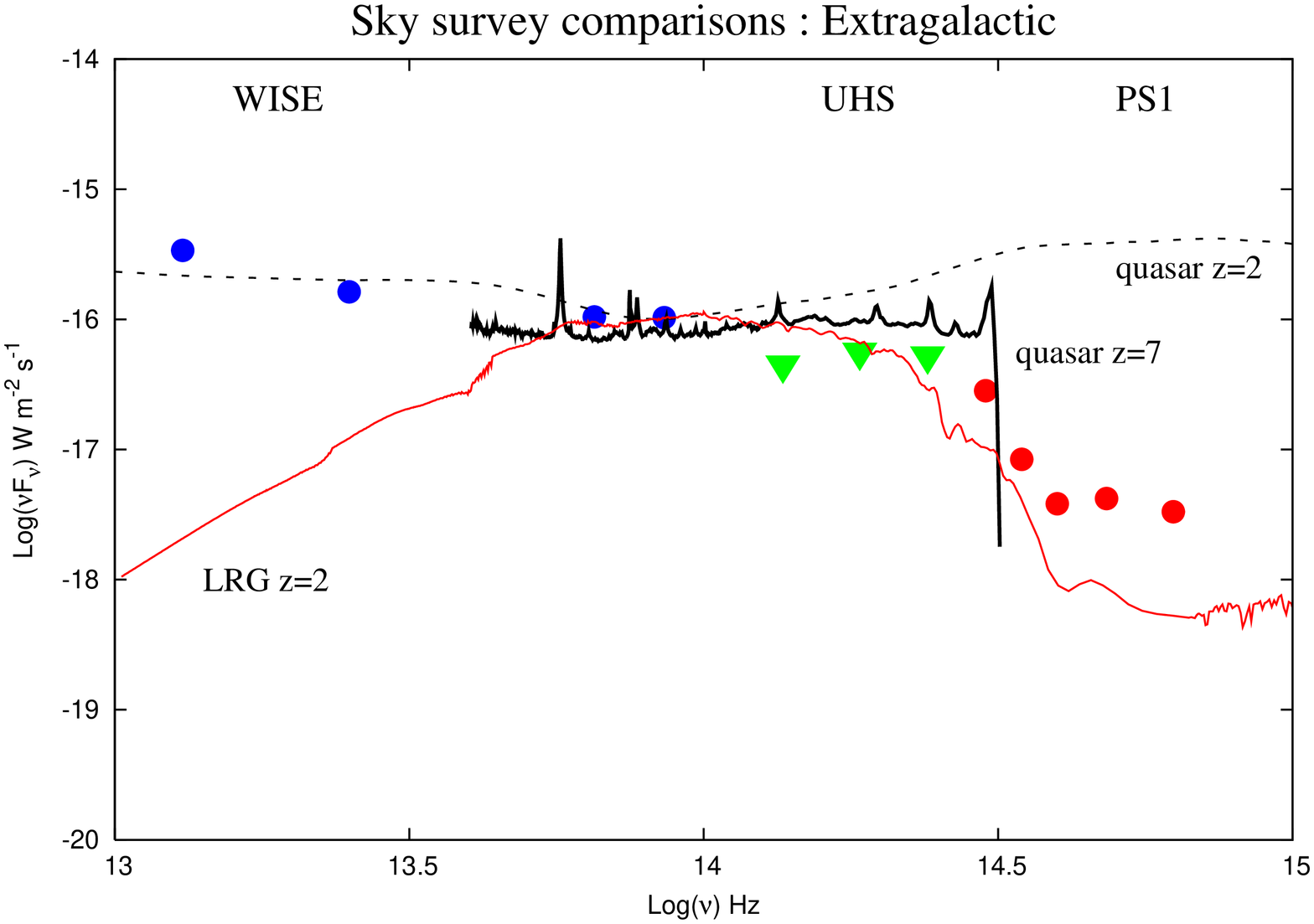}
\includegraphics[width=55mm,angle=-90]{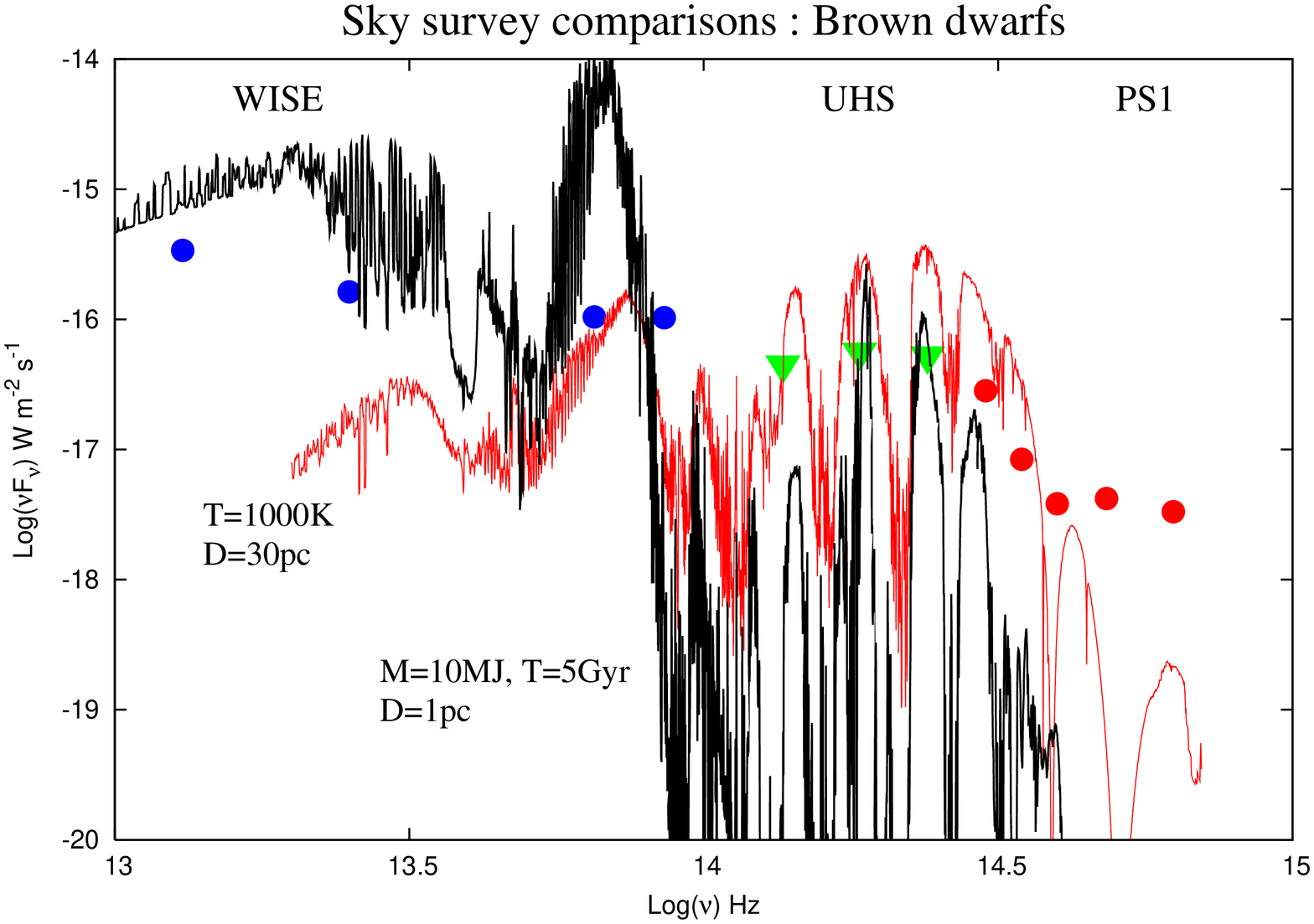}

\caption{\it Spectral energy distributions of various objects compared to 5$\sigma$ sensitivities of key sky surveys. Green triangles are JHK sensitivities for a proposed extension to the UKIDSS Large Area Survey, the UKIRT Hemisphere SUrvey (UHS).
Blue circles are for the WISE mission, from Mainzer et al 2005. Red circles (PS1) are for the PanSTARRS-1 3$\pi$ survey, taken from project documentation. The left hand frame compares extragalactic objects - a giant elliptical at $z=2$; the mean quasar continuum SED from Elvis et al 1994 redshifted to $z=2$; and a high redshift quasar spectrum redshifted to $z=7$. The right hand frame compares two model brown dwarf spectra, from Burrows et al (2003, 2006). The red line (lower curve at low frequency) is for an object with effective temperature of 1000K and surface gravity of 4.5, placed at a distance of 50pc. The black line (upper curve at low frequency) is for an object with mass of 10 Jupiter masses and age 5 Gyr, placed at a distance of 1pc. } 

\end{figure}

%

{\bf LSST}.

The Large Synoptic Survey Telescope (LSST) aims at the maximum possible \'{e}tendue, aiming at the same kind of science as PanSTARRS - hazardous NEOs, GRBs, supernovae, dark matter mapping via weak lensing - but a factor of several faster; it should be able to produce a survey equivalent to SDSS every few days. The design has an 8.4m telescope with a 10 sq.deg. field of view. The planned standard mode of use is to take 15 second exposures, and keep moving, covering the whole sky visible from LSST in bands $u g r i z y$ once every three days. This produces 15TB of imaging data every night. The aim is to keep up with this flow in quasi-real time, producing alerts for transient objects within minutes. This requires approximately 60 TFlops of processing power - a huge amount today, but following Moore's Law, very likely to be equivalent to merely the 500th most powerful computer in the world by 2012..  The LSST data management plan has a hierarchy of archive and data centres, reminiscent of the LHC Grid, with the primary mission facility acting like a `beamline', where a variety of research groups can rent space for their own experiments on the data flowing past. The LSST site has now been chosen (Cerro Pachon in Chile), but the project is not yet fully funded. More information can be found at http://www.lsst.org

\section{The end of survey discoveries ?}  \label{discovery}

In 1950, the universe seemed to consist of stars, and a sprinkling of dust. Over the last fifty years, the actual diverse and bizarre contents of the universe have been successively revealed as we surveyed the sky at a series of new wavelengths. Radio astronomy has shown us radio galaxies and pulsars; microwave observations have given us molecular clouds and the Big Bang fossil background; IR astronomy has shown us ultraluminous starburst galaxies and brown dwarfs; X-ray astronomy has given us collapsed object binaries and the intra-cluster medium; and submm astronomy has shown us debris disks and the epoch of galaxy formation. As well as revealing strange new objects, these surveys revealed new states of matter (relativistic plasma, degenerate matter, black holes) and new physical processes (bipolar ejection, matter-antimatter annihilation). Having opened up gamma-rays and the submm with GRO and SCUBA, there are no new wavelength windows left. Has this amazing journey of discovery now finished ? 

Wavelength is not the only possible axis of survey discovery space. Let us step through some other axes and examine their possibilities. In doing this, we will to some extent go over ground already trodden by Harwit (2003), but with a particular emphasis on surveys rather than discovery space in general, and with an eye to what is economically plausible.

{\bf Photon Flux.} Historically, going ever deeper has been as productive as opening new wavelength windows, the classic example of course being the existence of the entire extragalactic universe, which did not become apparent until reaching ten thousand times fainter than naked eye observations, requiring both large telescopes and the ability to integrate. We can now see things ten {\em billion} times fainter than the naked eye stars. However, we have reached the era of diminishing returns. The flux reached by a telescope is inversely proportional to diameter $D$ but its cost is proportional to $D^3$. Significant improvements can now only be achieved with world-scale facilities, and orders of magnitude improvements are unthinkable. The easy wins have been covered already - our detectors now achieve close to 100\% quantum efficiency; we have gone into space and reduced sky background to a minimum; and multi-night integrations have been used many times. We will keep building bigger telescopes, but it no longer seems the fast track to discovery. 

{\bf Spectral resolution.} Detailed spectroscopy of individual objects is of course the key technique of modern astrophysics. Spectroscopic surveys of samples drawn from imaging surveys have been carried out at many wavelengths, and have been particularly important for measuring redshift and so mapping the Universe in 3D; we were not expecting the voids, bubbles and walls that we found in the galaxy distribution in the 1980s. This industry will continue, but there is no obvious new barrier to break. Narrow band imaging surveys centred on specific atomic or molecular features (21cm HI, CO, H$\alpha$) have been fruitful, but again its not obvious there is anywhere new to go.

{\bf Polarization.} Polarisation measurements of individual objects are a very important physical diagnostic, but are polarisation surveys plausible ? Surveys of samples of known objects to the 0.1\% level have been done, with interesting results but no big surprises. Perhaps blank field imaging surveys in four Stokes parameters would turn up unexpected highly polarised objects ? This has essentially been done in radio astronomy but not at other wavelengths. 

{\bf Spatial resolution.} This is the dominant big-project target of the next few decades, and of course is the real point of Extremely Large Telescopes. Put together with multi-conjugate Adaptive Optics, we hope to achieve both depth and milli-arcsec resolution at the same time. However, the royal road to high spatial resolution is through interferometry. Surveys with radio interferometers in the twentieth century showed the existence of masers in space, and bulk relativistic outflow. In the twenty first century we will be doing microwave interferometry on the ground (ALMA) and IR interferometry in space (TPF/DARWIN), hoping to directly detect Earth-like planets around nearby stars. So there is excitement for at least some time to come; however, as with photon flux, we are hitting an economic brick wall. Significantly bigger and better experiments will be a very long time coming.

{\bf Time.} The observation of temporal changes has repeatedly brought about revolutionary changes in astronomy, the classic examples being Tycho's supernova, and the measurement of parallax. The last two decades has seen a renaissance in this area, with an impressive number of important discoveries from relatively cheap monitoring experiments - the discovery of extrasolar planets from velocity wobbles and transits; the discovery of the accelerating universe and dark energy from supernova campaigns; the location of substellar objects from survey proper motions; the existence of Trans-Neptunian Objects, and Near Earth Objects; the final pinning down of gamma-ray burst counterparts; and the limits on dark matter candidates from micro-lensing events. The next decade or two will see more ambitious photometric monitoring experiments, such as PanSTARRS and LSST, and a series of astrometric missions, culminating in GAIA, which will see external galaxies rotating. Overall, the `time window' is well and truly opened up. However, the temporal frequency axis is far from fully explored. My instinct is that this technique will continue to produce surprises for some time.
 
{\bf Non-light channels : particles.} Cosmic ray studies have been important for many decades, but you can't really do surveys - indeed the central mystery has alway been where they come from. Dark matter experiments are confronting what is arguably the most important problem in physics, let alone astrophysics, but again no survey is plausible. The big hope is neutrino astrophysics. Neutrinos should emerge from deep in the most fascinating places that we could otherwise never see - supernova cores, the centres of stars, the interior of quasar accretion discs. Measurement of solar neutrinos has solved a long standing problem, and set a challenge for particle physics - but what about the rest of the Universe ? New experiments such as ANTARES (under the sea) and AMANDA (under the ice) seem to be clearly detecting cosmic neutrinos, but no distinct sources have yet emerged. Possibly the next generation (ICECUBE) will get there. This looks like the best bet for genuinely unexpected discoveries in the twenty first century.

{\bf Non-light channels : gravitational waves.} Like neutrinos, we know that gravitational waves have to be there somewhere, and their existence has been indirectly proved by the famous binary pulsar timing experiment. However after many years of exquisite technical development, we still have no direct detection of a gravitational wave. The space interferometer mission LISA should finally detect gravitational waves, unless current predictions are badly wrong. However even LISA will not produce a genuine survey. We will detect many events and understand more astrophysics, but will have essentially no idea where they came from, except that hopefully some will correlate with Gamma-ray bursts. If we see totally unexpected signals, it will be very hard to know what to do next. 

{\bf Hyper-space planes : the Virtual Observatory.} As we explore the various possible axes one by one, many if not most of them are running out of steam, or too expensive to pursue. But we are a long way short of exploring the whole space - for example narrow line imaging in all Stokes parameters versus time. This exploration does not necessarily need complex new experiments. More survey-quality datasets come on line every year. As formats, access and query protocols, and analysis tool interaction protocols all get standardised, the virtual universe becomes easier for the e-astronomer to explore, and unexpected results will emerge. This, of course, is the agenda of the worldwide Virtual Observatory initiative.

\section{Conclusions}  \label{conclusions}

Surveys are perhaps the most cost effective and productive way of doing astronomy. In recent years optical, infra-red, and redshift surveys have produced spectacular results in determining cosmological parameters, finding the smallest stellar objects, decoding the history of the Milky Way, and much else besides. Surveys underway now and over the next few years should also produce impressive science. Having been the main engine of discovery for decades, there is a worry now that we have already explored every axis of discovery space. The best hopes for unexpected discoveries may be in massive time domain surveys, in neutrino astrophysics, and in exploring the full multi-dimensional space through the Virtual Observatory.

\newpage

\section{REFERENCES} \label{refs}

\noindent Becker, R.H. Fan, X., White, R.L.,  2001, AJ, 122, 2850.
\smallskip

\noindent Burrows, A., Sudarsky, D., Lurine, J.I., 2003 ApJ, 596, 587. 
\smallskip

\noindent Burrows, A., Sudarsky, D., Hubeny, I., 2006 ApJ, 640, 1063.
\smallskip

\noindent Cole, S., Percival, W.J., Peacock, J.A. \et 2005, MNRAS, 363 505.
\smallskip

\noindent Colless M.M., Dalton G.B., Maddox S.J., \et 2001, MNRAS, 328, 1039.
\smallskip

\noindent Dye, S., Warren, S.J., Hambly, N.C., \et 2006,  MNRAS, 372, 1227.
\smallskip

\noindent Eisenstein, D.J., Zehavi, I., Hogg, D.W., \et 2005, ApJ, 633, 560.
\smallskip

\noindent Efstathiou G., Moody S., Baugh C., \et 2002, MNRAS, 330, 29.
\smallskip

\noindent Elgarøy, Ø., Lahav, O., Percival, W.J., \et 2002, Phys.Rev.Lett., 89, 1301.
\smallskip

\noindent Ellis, S.C., Tinney, C.G., Burgasser, A.J., Kirkpatrick, J.D., McElwain, M.W.,
2005, AJ, 130, 2347.
\smallskip

\noindent Elvis, M, Wilkes, B.J., McDowell, J.C., Green, R.F., Bechtold, J., Willner, S.P., Oey, M.S., Polomski, E. Cutri, R., 1994, ApJSupp, 95, 1.	
\smallskip

\noindent Erdogdu, P., Huchra, J.P., Lahav, O., \et 2006, MNRAS, 368, 1515.
\smallskip

\noindent Fan, X., Narayanan, V.K., Lupton, R.H., \et 2001, AJ, 122, 2833.
\smallskip

\noindent Gunn, J.E., Peterson, B.A., 1965, ApJ, 142, 1633.
\smallskip

\noindent Harwit, M., 2003, Physics Today, November 2003, 38.
\smallskip

\noindent Hewett, P.C., Warren, S.J., Leggett, S.K., Hodgkin, S.L., 2006 MNRAS, 367, 454.
\smallskip

\noindent Huetsi, G., 2005, A\&A, 449, 891.
\smallskip

\noindent Ibata, R.A., Gilmore, G., Irwin, M.I. 1994 Nature, 370, 194.
\smallskip

\noindent Jarrett, T.H. 2004, PASA, 21, 396.
\smallskip

\noindent Jones, D.H., Saunders, W., Colless, M., \et 2004, MNRAS, 355, 747.
\smallskip

\noindent Juric, M., Ivezic, Z., Brooks, A., \et 2005, ApJ submitted (astro-ph/0510520)
\smallskip

\noindent Kirkpatrick, J.D., Reid, I.N., Liebert, J., Cutri, R.M., Nelson, B., Beichman, C.A., Dahn, C.C., Monet, D.G., Gizis, J.E., Skrutskie, M.F., 1999, ApJ, 519, 802.
\smallskip

\noindent Kogut, A., Spergel, D.N., Barnes, C., \et 2003 ApJSupp, 148, 161.
\smallskip

\noindent Lawrence, A., Warren, S.J., Almaini, O., \et\ 2006 MNRAS submitted (astro-ph/0604426) 
\smallskip

\noindent Lodieu, N., Hambly, N.C., Jameson, R.F., Hodgkin, S.T., Carraro, G., 
Kendall, T.R., 2007, MNRAS, 374, 372.
\smallskip

\noindent Maddox, S.J., Sutherland, W.J., Efstathiou, G., Loveday, J., 1990, MNRAS, 243, 692.
\smallskip

\noindent 
Mainzer, A.K., Eisenhardt, P., Wright, E.L., Liu, F-C., Irace, W., Heinrichsen, I., Cutri, R., Duval, V., 2006, Proc SPIE, 6256, 61.
\smallskip

\noindent Majewski, S.R., Skrutskie, M.F., Weinberg, M.D., Ostheimer, J.C.	
2003, 599, 1082.
\smallskip

\noindent McPherson, A.M., Born, A.,  Sutherland, S., Emerson, J., Little, B., Jeffers, P.,  Stewart, M., Murray, J., Ward, K., 2006, Proc SPIE 6267, 7.
\smallskip

\noindent Nakajima, T., Oppenheimer, B.R., Kulkarni, S.R., Golimowski, D.A., Matthews, K., Durrance, S.T., 1995 Nature 378 463.
\smallskip

\noindent Percival W.J., Baugh C.M., Bland-Hawthorn J., \et 2001, MNRAS, 327, 1297.
\smallskip

\noindent Percival W.J., Sutherland W.J., Peacock J.A., \et 2002, MNRAS 337, 1068 
\smallskip

\noindent Peacock J.A., 2002, ASP Conf Series, 283, 19.
\smallskip

\noindent Pope, Adrian C.; Matsubara, Takahiko; Szalay, Alexander S., \et 2004, ApJ, 607, 655.
\smallskip

\noindent Saunders, W., Sutherland, W.J., Maddox, S.J., \et 2000, MNRAS, 317,55.
\smallskip

\noindent Skrutskie M.F., Cutri R.M., Stiening, \et 2006, AJ, 131, 1163.
\smallskip

\noindent Tegmark, M., Blanton, M., Strauss, M., \et 2004, ApJ, 606, 702.
\smallskip

\noindent Warren, S.J., Hambly, N.C., Dye, S., \et 2007a, MNRAS in press (astro-ph/0610191)
\smallskip

\noindent Warren, S.J., Mortlock, D.J., Legget, S.K., \et 2007b, MNRAS submitted
\smallskip

\noindent White, R.L., Becker, R.H., Fan, X., Strauss, M.A., 2003 AJ, 126, 1.
\smallskip

\noindent York, D.G., Adelman, J., Anderson, J.E., \et 2000, AJ, 120, 1579
\smallskip

\end{document}